\documentclass{aa}

\usepackage{amsmath}
\usepackage{epsfig}
\usepackage{graphicx}
\usepackage{lscape}
\usepackage{subfig}
\usepackage{natbib}
\usepackage{txfonts}

\usepackage{amssymb}
\usepackage{color}
\usepackage{url}
\usepackage{xspace}
\usepackage[flushleft]{threeparttable}
\usepackage{hyperref}
\hypersetup{
     citecolor=blue,
     colorlinks=true,
     linkcolor=blue,
     urlcolor=magenta,
}

\newcommand {\bc}{\begin {center}}
\newcommand {\ec}{\end {center}}
\newcommand {\be}{\begin {equation}}
\newcommand {\ee}{\end {equation}}
\newcommand {\beq}{\begin {eqnarray}}
\newcommand {\eeq}{\end {eqnarray}}

\def\flux{erg~s$^{-1}$~cm$^{-2}$\xspace}
\def\lum{erg~s$^{-1}$\xspace}
\def\xmm{{\it XMM-Newton}\xspace}
\def\nustar{{\it NuSTAR}\xspace}

\usepackage[colorinlistoftodos]{todonotes}

\def\src{SXP\,1062\xspace}

\begin{document}

\title{The unusual behaviour of the young X-ray pulsar SXP\,1062 during the 2019 outburst}

  \author{Sergey~S.~Tsygankov \inst{1,2}
\and  Victor Doroshenko \inst{3,2}
\and  Alexander A.~Mushtukov \inst{4,2,5}
\and  Frank Haberl \inst{6}
\and  Georgios~Vasilopoulos \inst{7}
\and  Chandreyee Maitra \inst{6}
\and  Andrea Santangelo \inst{3}
\and  Alexander A. Lutovinov \inst{2}
\and  Juri Poutanen \inst{1,2,8}
  }

   \institute{Department of Physics and Astronomy, FI-20014 University of Turku, Finland;  
              \email{sergey.tsygankov@utu.fi}
       \and
             Space Research Institute of the Russian Academy of Sciences, Profsoyuznaya Str. 84/32, Moscow 117997, Russia
       \and
              Institut f\"ur Astronomie und Astrophysik, Universit\"at T\"ubingen, Sand 1, D-72076 T\"ubingen, Germany
       \and
              Leiden Observatory, Leiden University, NL-2300RA Leiden, The Netherlands
       \and
             Pulkovo Observatory, Russian Academy of Sciences, Saint Petersburg 196140, Russia
       \and
             Max-Planck-Institut f{\"u}r extraterrestrische Physik, Gie{\ss}enbachstra{\ss}e, 85748 Garching, Germany
       \and
            Department of Astronomy, Yale University, PO Box 208101, New Haven, CT 06520-8101, USA
       \and
           Nordita, KTH Royal Institute of Technology and Stockholm University, Roslagstullsbacken 23, SE-10691 Stockholm, Sweden
          }
   \titlerunning{SXP\,1062 during 2019 outburst}
   \authorrunning{S. Tsygankov et al. }
   \date{Received 16.01.2020; accepted 24.03.2020}

\abstract
{We present results of the first dedicated observation of the young X-ray pulsar \src in the broad X-ray energy band obtained during its 2019 outburst with the \nustar and \xmm observatories. The analysis of the pulse-phase averaged and phase-resolved spectra in the energy band from 0.5 to 70 keV did not reveal any  evidence for the presence of a cyclotron line. The spin period of the pulsar was found to have decreased to $979.48\pm0.06$~s implying a $\sim$10\% reduction compared to the last measured period during the monitoring campaign conducted about five years ago, and is puzzling considering that the system apparently did not show major outbursts ever since.
The switch of the pulsar to the spin-up regime supports the common assumption that torques acting on the accreting neutron star are nearly balanced and thus \src likely also spins with a period close to the equilibrium value for this system. The current monitoring of the source revealed also a sharp drop of its soft X-ray flux right after the outburst, which is in drastic contrast to the behaviour during the previous outburst when the pulsar remained observable for years with only a minor flux decrease after the end of the outburst. This unexpected off state of the source lasted for at most 20 days after which \src returned to the level observed during previous campaigns. 
We discuss this and other findings in context of the modern models of accretion onto strongly magnetized neutron stars.
}


\maketitle

\section{Introduction}
\label{intro}

The existence of neutron stars (NSs) with extremely strong magnetic fields in binary systems is a stumbling-stone of modern astrophysics. From one side, it appears to contradict the standard theory of magnetic field decay \citep[see e.g.,][]{2000ApJ...529L..29C}. From another, \cite{2018MNRAS.473.3204I} demonstrated that under certain conditions a very strong magnetic field may survive allowing the existence of such sources with ages of a few Myr or even older. Observed long spin periods and spin evolution of many systems including \src \citep{1975SvAL....1..223S,2010A&A...515A..10D,2013PhyU...56..321S} also point to ultra-strong fields in many cases, although ambiguities in theoretical modeling of the torques make it hard to draw definite conclusions here. The detailed investigation of such long spin period pulsars, especially with known age, would therefore be very important and could potentially shed new light on the physics of magnetic field evolution and on the origin of
ultra-luminous X-ray pulsars which can be powered by the accretion onto NSs with extremely strong magnetic field \citep[see e.g.][]{1975AA....42..311B, 2015MNRAS.454.2539M}.

One of the most promising sources in this respect is \src\ located in the Small Magellanic Cloud (SMC), which is one of the longest-period X-ray pulsars (XRPs) known to date. It was discovered by \cite{2012MNRAS.420L..13H} and appears to be associated with the supernova remnant (SNR) MCSNR J0127$-$7332
\citep{2012A&A...537L...1H}. \cite{2012MNRAS.420L..13H} established the spectral class of the optical companion in the system as a Be star. The association of the pulsar with the SNR allowed to constrain its age to be 10-40 kyr \citep{2012MNRAS.420L..13H,2012A&A...537L...1H}, making the source a very rare case among high-mass X-ray binaries \citep[for a list of SNR associations see][]{2019MNRAS.490.5494M}. Despite such a young age the pulsar has a very long spin period of $P=1062$~s and continued to spin down at least until 2014 when the period reached a value around 1080~s \citep{2013A&A...556A.139S,2018MNRAS.475.2809G}.

The origin of such a long period is a matter of intense debates in the literature and is relevant in the context of possible distribution of birth periods and magnetic fields of NSs. Despite the disagreement on details between different models, most of the authors claim that an 
above-average magnetic field for the neutron star with $B\sim10^{13-15}$~G is required to explain properties of the system \citep{2012MNRAS.421L.127P,2012ApJ...757..171F,2012MNRAS.424L..39I,2013PhyU...56..321S,2017MNRAS.471.4982S}. Otherwise, an unusually long spin period (at least 0.5~s) at the NS birth was suggested by \cite{2012A&A...537L...1H}.
No evidence for a cyclotron resonant scattering feature (CRSF) in the soft X-ray band was found in the \xmm data \citep{2013A&A...556A.139S}. The search for cyclotron lines above 10 keV is performed in the current work for the first time and only became possible with the launch of the \nustar observatory with its broadband coverage, high sensitivity and good angular resolution.

The very long spin period of \src makes its accurate measurement a very challenging task owing to relatively short available observations (mainly with the {\it Swift}/XRT telescope). As a result, different authors provide different interpretations of the same data sets. For instance, using a two-year long monitoring of \src with different X-ray observatories,  \cite{2017MNRAS.471.4982S} found an evidence for a glitch event, which happened about 25 days after the peak of the Type I outburst in the middle of 2014, and even the glitch magnitude was found to be consistent with values predicted by \citet{2015A&A...578A..52D} for accreting pulsars. At the same time the same data set was interpreted as ordinary spin-up of the NS due to an enhanced accretion rate \citep{2018MNRAS.475.2809G}. Therefore, the timing analysis of the data obtained during the  current 2019 outburst is essential to continue monitoring the evolution of the spin period, which was one of the main goals of our work. 

An almost two-year long orbital period of the binary system \citep[$P_{\rm orb}=656.5\pm0.5$~d; ][]{2012ATel.4596....1S,2018MNRAS.475.2809G,2019ATel12890....1S} and its low flux in quiescence require extremely time-consuming monitoring programs in order to investigate the  behaviour of the source during the entire orbital cycle. However, such observations were performed with the {\it Swift}/XRT telescope between October 2012 and November 2014 \citep[see e.g., Fig. 1 in ][]{2018MNRAS.475.2809G}. Particularly, it was found that during the periastron passage of the neutron star, \src exhibited a bright outburst with the peak luminosity around $5\times10^{37}$~\lum, followed by a decrease down to $\sim6\times10^{36}$~\lum in $\sim$100 days. After that, the observed flux essentially leveled out, which made the source observable during a complete orbital cycle. The minimal luminosity reached by the pulsar just before the next Type I outburst was $\sim1\times10^{36}$~\lum.
According to the model proposed by \cite{2017A&A...608A..17T}, such long-term behaviour is typical for slowly rotating XRPs owing to their transition to the stable accretion from the weakly-ionized (``cold'') disc below some critical luminosity $L_{\rm cold}$. It is worth noting that this luminosity depends on the inner radius of the accretion disc and, hence, can be used as indirect estimate of the strength of the NS magnetic field.

Here we present the results of the first dedicated observation of \src in the broad energy band with the \nustar and \xmm observatories. The long exposure allowed us to perform  accurate timing and spectral analysis of the source. The obtained results are discussed in the context of the physical models of accretion onto strongly magnetized NSs.

\section{Data analysis}
\label{sec:data}

\src exhibits a very rich phenomenology on different time scales. Therefore, for our research we used archival data as well as observations obtained during the monitoring program initiated by us to study the properties of the source around the 2019 outburst.

The campaign started on 2019 October 5 (MJD 58761.13), with observations using XRT \citep{2005SSRv..120..165B} on-board the {\it Neil Gehrels Swift Observatory} \citep{2004ApJ...611.1005G}. The main goal of this monitoring was to register the transition of \src into the outburst and to trigger a target of opportunity \nustar observation. The source has been found in a bright state on October 22 (MJD 58778.76; XRT ObsID 00048719112). The next XRT observation (ObsID 00048719113) was performed simultaneously with \nustar on October 25 (MJD 58781.85). All XRT data were taken in the photon counting mode. In the observations where the source was significantly detected the energy flux was estimated by fitting respective spectra extracted using the online tools\footnote{\url{http://www.swift.ac.uk/user_objects/}} \citep[][]{2009MNRAS.397.1177E} provided by the UK Swift Science Data Centre. Images of the field were also obtained with the same tools.

The \textit{NuSTAR} \citep{2013ApJ...770..103H} observation (ObsID 90501344002) with an effective exposure about 40 ks (with total elapsed time of $\sim$83 ks) has been performed right after the peak of the outburst when \src had a luminosity of $\sim2.5\times10^{37}$~\lum, assuming the distance to the source of $d=60$~kpc \citep[this value is used throughout the paper; see][]{2014ApJ...780...59G}. 
The raw data were reduced following the standard procedures described in the \nustar user guide, and using the standard \nustar Data Analysis Software {\sc nustardas} v1.8.0 and the CALDB version 20191008.  
The source and the background counts were extracted from circular regions with radii of 50\arcsec\ and 100\arcsec, respectively, using the {\sc nuproducts} routine. The background region was located in the corner of the same chip. 

\begin{figure}
\centering 
\includegraphics[width=0.9\columnwidth, bb=50 190 570 695]{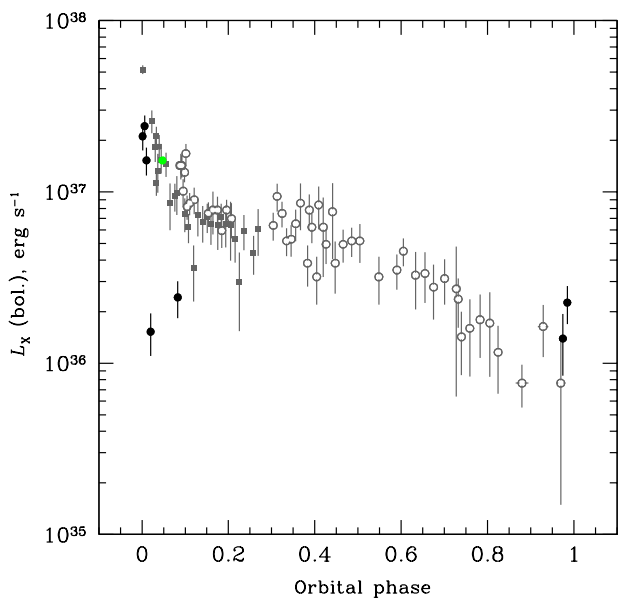}
\includegraphics[width=0.9\columnwidth, bb=50 190 570 695]{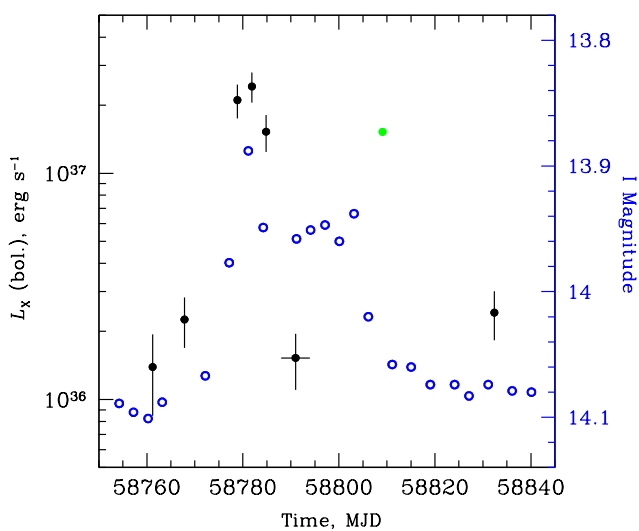}
\caption{{\it Top}: 
Bolometrically corrected orbital light curve of \src obtained in 2019 with {\it Swift}/XRT (black filled circles) and \xmm (green circle). The original measurements are folded using ephemeris from \cite{2019ATel12890....1S}. Grey open circles and filled squares represent the data obtained during two consecutive orbital cycles covering the period from October 2012 to November 2014. 
{\it Bottom}: Zoomed part of the light curve obtained in 2019. The blue open points represent the OGLE data in the $I$ band. 
}
\label{fig:lcxrt}
\end{figure}

The data from {\it Swift}/XRT and \nustar were used in the 0.3--10 and 3--79~keV bands, respectively. 
To fit the spectra we used the {\sc xspec} package \citep{Arn96}. All spectra were binned to have at least 1 count per energy bin and W-statistic\footnote{\url{https://heasarc.gsfc.nasa.gov/xanadu/xspec/manual/XSappendixStatistics.html}} was applied
\citep{1979ApJ...230..274W}. The spectra from the two \nustar telescopes and the simultaneous {\it Swift}/XRT observation were modeled together, allowing a free cross-normalization constant to account for possible differences in absolute flux calibration of the individual telescopes. To be able to compare the results of our monitoring with the archival XRT data, all of those were reduced in the same manner.

Continuing monitoring of the source with the XRT telescope revealed an abrupt drop of the flux starting from MJD 58788, just three days after the previous detection in the bright state with luminosity slightly above $10^{37}$~\lum (ObsID 00048719114). This low-flux interval was covered by three consecutive XRT observations (ObsIDs 00048719115-17). Unfortunately, the exposure of individual pointings (around 1 ks) turned out to be insufficient to detect the source or provide a constraining upper limit on the source flux (typical 3$\sigma$ upper limit is just $\sim$5 times less than the flux measured during observation 00048719114). Therefore, we averaged all three observations that allowed us to increase the total exposure up to 3.7 ks. Analysis of the resulting image in 0.3--10 keV range revealed 14 counts inside the aperture with 40\arcsec\ radius centered at the source position. To estimate the expected background count rate we chose the much larger area defined by the annulus centered at the source position with the inner and outer radii of 70\arcsec\ and 250\arcsec, respectively. As a result we detected 100 counts that can be rescaled to slightly less than 3 counts expected within the source aperture. According to the Poisson statistics, the detected 14 counts with expected background of 3 counts correspond to the detection of the source at $3\sigma$ significance level \citep[see e.g.,][]{1986ApJ...303..336G}. The same procedure resulted in 62 net counts collected during 1.7 ks exposure in the last bright observation 00048719114. Assuming the same spectral shape in both states, the flux level in the low state was then determined as $F_{0.3-10}=8.5\times10^{-13}$~\flux from rescaling of the corresponding count rates, i.e. a factor of ten lower than during the previous observation.

After the flux drop from the source was discovered in the {\it Swift}/XRT data we requested a DDT observation with \xmm. 
The requested observations (ObsID 0853980901) were performed on 2019 November 21-22 (MJD 58808.88333-58809.28738). EPIC pn and EPIC MOS detectors were operated in the full frame mode with thin filters. The effective exposures for the three main detectors are 17.7 ks (pn), 23.2 ks (MOS1) and 23.7 (MOS2). The data reduction procedures using the latest version of the \xmm\  {\sc Science Analysis Software} (SAS; version 18.0) were applied following standard procedures.\footnote{\url{https://www.cosmos.esa.int/web/xmm-newton/sas-threads}} Particularly, the data were grade filtered using patterns 0-12 and 0-4 for MOS and pn data, respectively, and {\sc FLAG==0} option applied.
Circular regions with radii of 15\arcsec\ and 55\arcsec\ were chosen to extract the source and the background photons, respectively. The background region was placed close to the source at the same CCD and avoiding other point sources. We also checked the data for the presence of soft proton background flares, but detected none. Ancillary response files (arf) were generated for each spectrum using {\sc arfgen} and  redistribution matrices were generated using {\sc rmfgen}.

\section{Results}

In order to ensure a detection of \src near the maximum of the Type I outburst we initiated the {\it Swift}/XRT monitoring around orbital phase 0.97 \citep[according to the ephemeris derived by][]{2019ATel12890....1S}. The first two pointings revealed the source in a low-luminosity state around $1\times10^{36}$~\lum, as was expected from extrapolation of the source behaviour observed during periastron passages in 2012 and 2014 (Fig.~\ref{fig:lcxrt}). The next observation revealed the source in the bright state and triggered the first dedicated observation in hard X-rays with the \nustar observatory. The most conspicuous feature of the light curve presented in Fig.~\ref{fig:lcxrt} is a sharp drop of the flux by  about a factor of 10, which happened soon after the outburst peak. Interestingly, the luminosity in this low state is roughly the same as the lowest luminosity observed from the source in the very end of the previous orbital cycle. The following \xmm observation performed about 20 days later (MJD 58809) revealed the source at roughly the same luminosity as before the flux drop (shown with the green point). The next observation 23 days later (MJD 58832) had been done with the {\it Swift}/XRT telescope and found \src again with a flux significantly lower than one may expect based on the previous monitoring of the source in 2012--2014. Interestingly, the observed decrease of the X-ray flux coincides with an also unusual behaviour of the source light curve in the $I$ band \citep[][]{2020ATel13426....1S}. Both light curves in the $I$ band (OGLE data) and in the X-rays are presented at the bottom panel of Fig.~\ref{fig:lcxrt}.\footnote{\url{http://ogle.astrouw.edu.pl/ogle4/xrom/xrom.html}; \cite{2008AcA....58..187U}}

\subsection{Timing analysis}

\begin{figure}
\centering
\includegraphics[width=0.9\columnwidth, bb=30 390 520 680]{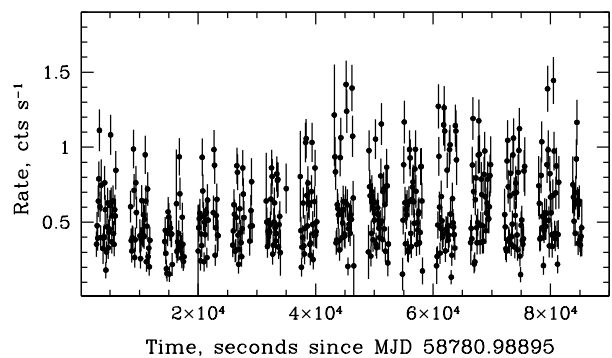}
\caption{Light curve of \src obtained with \nustar in the 3--79 keV energy range. The total count rate for both modules is shown with 100~s time binning. }
 \label{fig:lcnu}
\end{figure}

The light curve of \src in the 3--79 keV energy band based on the \nustar data is shown in Fig.~\ref{fig:lcnu}. No strong variability except arising from the pulsations from the source are seen. The averaged count rate after background subtraction is $\sim$0.35 cnt s$^{-1}$ in each module. The long exposure and comparatively high flux level allowed to easily see individual pulses and determine the spin period of the source as $P_{\rm s}=979.50(7)$~s using the phase-connection technique \citep{1981ApJ...247.1003D}. Here we used the average pulse profile of the source found by folding of the co-added light curve of the two \nustar units in the 3--79\,keV energy band with initial period determined using epoch folding as a template, and then determined the time of arrival for each of 56 individual pulses by fitting this template to the light curve.  The period was then determined by fitting the times of arrival with a constant period model, i.e. $t_{\rm n}=n\times P_{\rm spin}$. 
Considering that folded template has also some uncertainty, this was added as systematic error to individual light-curve points when performing the fit. To account for possible systematic effects associated with the distortion of the template due to error in the initially assumed period, we repeated the entire procedure 100 times assuming periods between 978 and 982\,s to obtain the template. The final reported period value and uncertainty are then the mean and the standard deviation of the best-fit values within this sample.
No evidence for spin-up was found, which is, however, not surprising given the short duration of the \nustar observation and estimated uncertainties for times of arrival for individual pulses (on average $\sim$17\,s).
We note that the final period value is about 100~s shorter than was measured during the last long-term monitoring of the source about five years ago \citep{2018MNRAS.475.2809G}. 

Using the \xmm data collected 27 days later and following the same techniques, we were able to confirm the long-term spin-up trend discovered with the \nustar data. At the same time, the  measured period $P_{\rm s}=981.8(2)$ (estimated using the same procedure) is somewhat longer than that derived from the \nustar observation, implying an average spin-down between the two observations at the level of about $-0.085(8)$ s d$^{-1}$. The observed spin-down is actually not surprising as the source was apparently accreting at significantly lower rate for at least one week between the two period measurements (see below).

\begin{figure}
\centering
\includegraphics[width=0.9\columnwidth, bb=65 340 500 710]{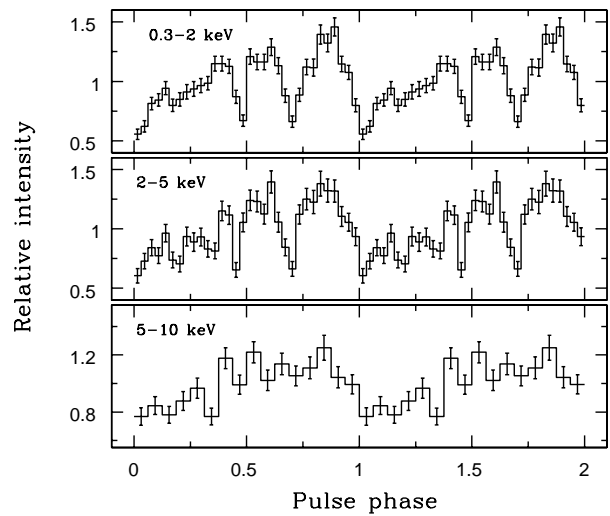}
\caption{Pulse profile of \src in different energy bands from the \xmm observation at luminosity $L_{\rm bol}=1.4\times10^{37}$~\lum.}
 \label{fig:pprofxmm}
\end{figure}
\begin{figure}
\centering
\includegraphics[width=0.9\columnwidth, bb=65 235 500 710]{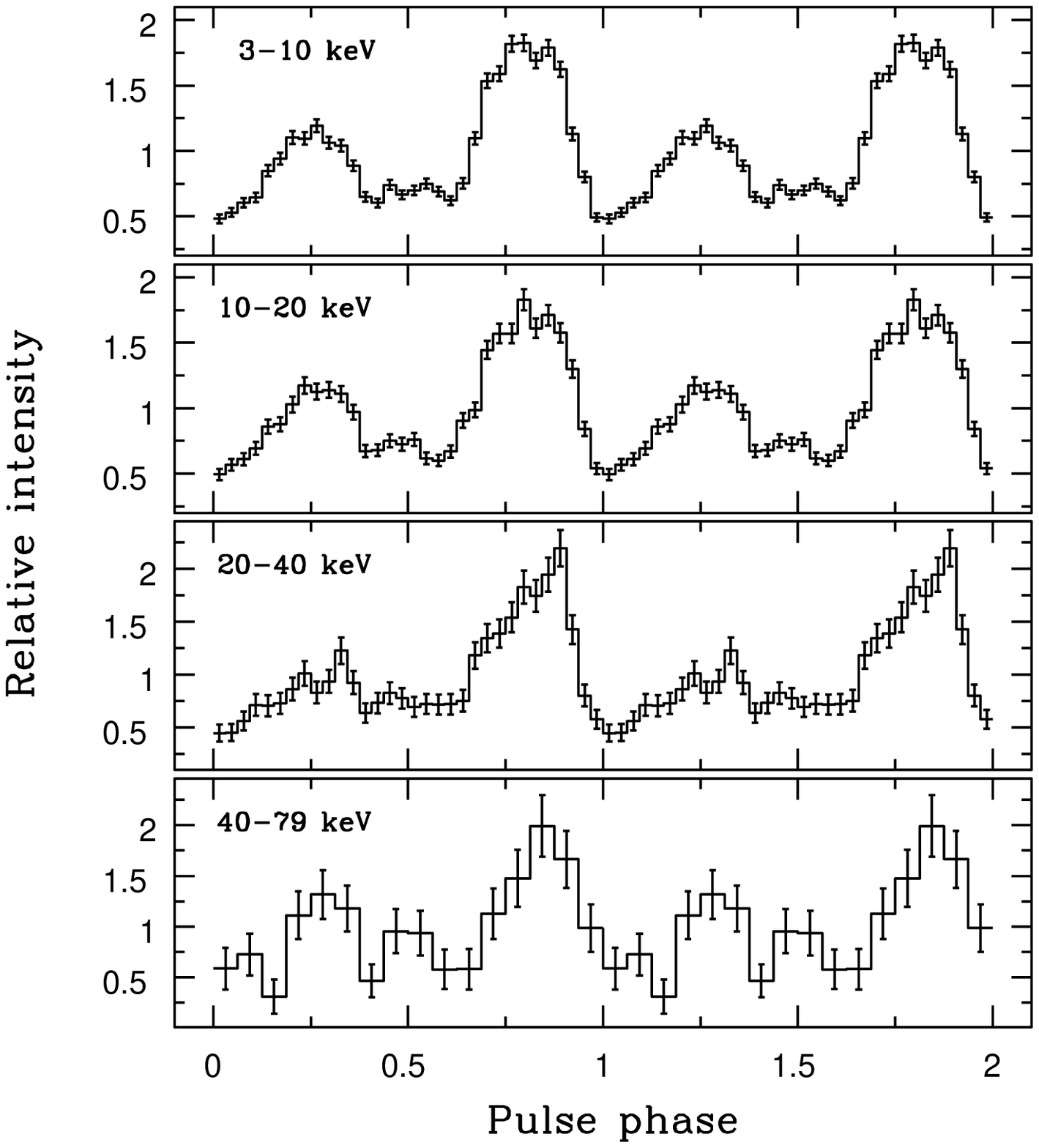}
\caption{Pulse profile of \src in different energy bands from the \nustar observation at luminosity $L_{\rm bol}=2.4\times10^{37}$~\lum.}
 \label{fig:pprofnu}
\end{figure}
Folding the X-ray light curves of the source obtained with \xmm and \nustar in different energy bands with the corresponding spin periods we were able to study the evolution of the pulse profile shape as a function of energy (Figs.~\ref{fig:pprofxmm} and \ref{fig:pprofnu}). As can be seen, the pulse profiles during the two observations (and in different energy bands) are quite different. In the \xmm data the pulse profiles are complex consisting of three broad peaks, whereas at higher energies they are very similar in four energy bands from 3 to 79 keV and consist of two main peaks and a plateau around phase 0.5, that can be considered as third, much weaker peak (which was stronger during the \xmm observation / at lower energies). Pulsed fraction determined as $\mathrm{PF}=(F_\mathrm{max}-F_\mathrm{min})/(F_\mathrm{max}+F_\mathrm{min})$ (where $F_\mathrm{max}$ and $F_\mathrm{min}$ are maximum and
minimum fluxes in the pulse profile, respectively) is quite high $\sim$40\% in the \xmm bands and $\sim$60\% in the \nustar data and does not appear to depend on energy during given observation.

\subsection{Spectral analysis}

The spectral properties of \src are well studied in the standard X-ray band below $\sim$10\,keV. Particularly, it was shown that the spectrum can be described by the standard absorbed power law with the possible presence of a soft excess below 1 keV and an iron $K_{\rm \alpha}$ emission line at 6.4 keV \citep[see e.g.,][]{2012MNRAS.420L..13H,2013A&A...556A.139S}. Taking into account multiple hints for an ultra-strong magnetic field in \src, one could expect the presence of either electron or proton CRSF in the source spectrum above 10~keV.

\begin{figure}
\centering
\includegraphics[width=0.9\columnwidth, bb=40 275 550 695]{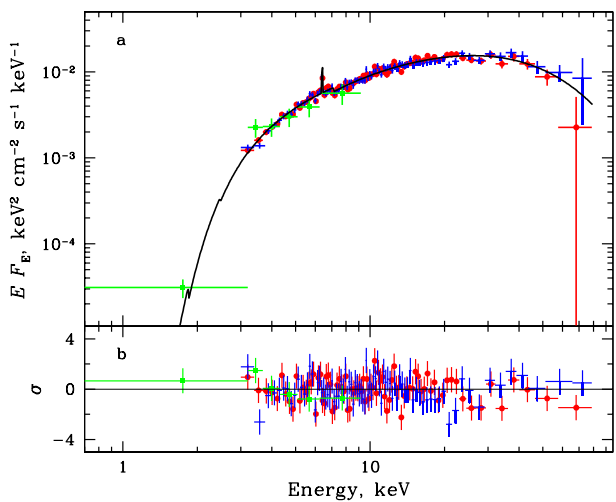}
\caption{(a) Broad-band spectrum of \src from simultaneous {\it Swift}/XRT and \nustar observations, and (b) corresponding residuals from the best-fit model. 
Green, red and blue crosses represent data obtained from {\it Swift}/XRT, \nustar-FPMA/B, respectively. The solid line shows the best-fit model consisting of absorbed power law with high-energy cutoff and iron emission line.}
 \label{fig:broadspec}
\end{figure}

The broad-band spectrum obtained with the {\it Swift}/XRT and \nustar instruments turned out to be typical of XRPs and is presented in Fig.~\ref{fig:broadspec}. The continuum can be modeled with either phenomenological models (like power law with high-energy exponential cutoff, {\sc cutoffpl} in {\sc xspec}) or physically motivated (e.g. thermal Comptonization, {\sc thcomp} in {\sc xspec}) models. In our analysis we use the more commonly applied {\sc cutoffpl} model, which also has the least number of free parameters.

The continuum was modified with a Gaussian emission line at 6.4~keV to account for the iron line and with photo-electric absorption at low energies ({\sc phabs} in {\sc xspec}). To take into account possible differences in absolute normalization in the two \nustar modules and the {\it Swift}/XRT telescope, cross-normalization constants were introduced to the model. The spectrum can be described well with this simple model and no evidence for any narrow feature which could be associated with a putative cyclotron line was found. The resulting best-fit parameters are:
$N_{\rm H}=(7.4\pm0.8)\times10^{22}$~cm$^{-2}$, photon index $\Gamma=0.63\pm0.06$, and e-folding energy $E_{\rm cut}=18.8\pm1.2$~keV. Position and width of the iron line were frozen at 6.4 keV and 0.01 keV, respectively, resulting in a line flux of $(1.1\pm0.3)\times10^{-5}$~\flux. The C-statistic value for a given approximation is 2052 for 2127 degrees of freedom.
The unabsorbed broad-band flux $F_{\rm 0.3-100 keV}=(5.6\pm0.1)\times10^{-11}$~\flux corresponds to a luminosity $L=(2.41\pm0.05)\times10^{37}$~\lum.

The observed broad-band spectral model was used to determine the bolometric correction $K_{\rm bol}=3.9$ required in order to convert the fluxes measured with {\it Swift}/XRT in the 0.5--10 keV range to the broad 0.5--100 keV range.
In our study we assume that this correction factor does not depend on the source luminosity, which appears justified considering the relative stability of the source spectrum in the soft X-ray band \citep{2018MNRAS.475.2809G}. However, this assumption should be taken with caution. Because no broadband spectral data are available at low fluxes, the actual spectrum of \src  can be slightly softer there, that may result in a smaller bolometric correction factor and, hence, lower luminosity.

Because the simultaneous {\it Swift} observations were too short, we have performed a pulse-phase resolved spectral analysis using only \nustar data and the model described above. The absorption column density and the iron line energy and width were fixed to the values derived from the phase-averaged spectrum. The spectra at different pulse phases do not show strong variations, as is also evident from the similarity of the pulse profiles in different energy bands. No features possibly associated with a CRSF were found in the phase-resolved spectra either.

A study of the spectral evolution of the source during the periastron passage in 2014 by \citet{2018MNRAS.475.2809G} revealed that soon after the peak of the outburst the absorption column density had significantly increased for a short time. The authors interpreted this as obscuration of the NS by the Be star disc material. Our current monitoring with the XRT telescope confirms a sharp increase of the $N_{\rm H}$ value from being consistent with zero to  $(6.8\pm2.5)\times10^{22}$~cm$^{-2}$ between MJD 58778.76 (ObsID 00048719112) and 58781.85 (ObsID 00048719113), assuming a simple absorbed power-law model.
In the following observation the absorption column density was found to keep low value.

One month after the \nustar pointing, we observed \src with \xmm. The EPIC spectra in the 0.5-10\,keV band are well described with a combination of the power-law and blackbody models modified by photo-electric absorption ({\sc phabs(bb+pow)} in {\sc xspec}). The resulting best-fit parameters are very similar to the results obtained earlier using \xmm data \citep[e.g.,][]{2013A&A...556A.139S}: $N_{\rm H}=(0.22\pm0.02)\times10^{22}$~cm$^{-2}$, the photon index $\Gamma=0.82\pm0.02$, and the blackbody temperature $kT_{\rm bb}=0.20\pm0.01$~keV with a C-statistic value of 4525 for 4752 degrees of freedom. Because the EPIC pn, MOS1 and MOS2 data were fitted simultaneously, cross-normalization factors were introduced to the model with resulting values consistent with unity within 2--3\%. The unabsorbed bolometrically-corrected flux from \src in this observation is $(3.32\pm0.05)\times10^{-11}$~\flux that corresponds to a luminosity $L=(1.43\pm0.02)\times10^{37}$~\lum, i.e. comparable to the \nustar observation and by an order of magnitude higher than the value obtained from the combination of three preceding XRT observations.

\begin{figure}
\centering
\includegraphics[width=0.9\columnwidth, bb=40 275 550 695]{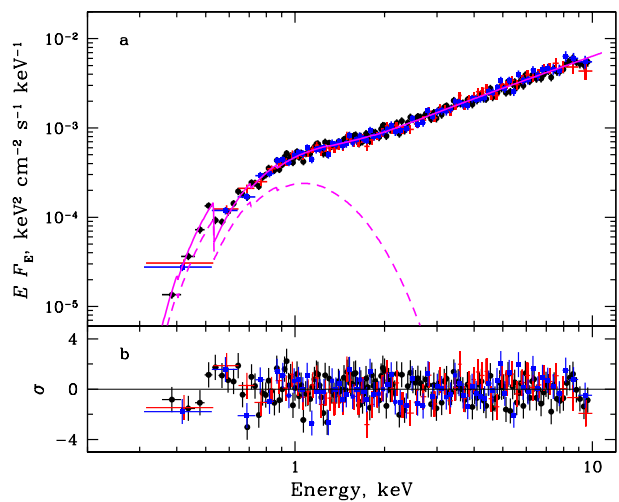}
\caption{\xmm spectrum of \src obtained on MJD 58809 after its re-brightening. The EPIC pn, MOS1 and MOS2 data are shown with black, red and blue points, respectively. The solid magenta curve corresponds to the best-fit model consisting of the blackbody (shown with the dashed curve) and absorbed power-law components. }
 \label{fig:xmmspec}
\end{figure}

\section{Discussion and conclusions}
\label{sec:discus}

We presented here the results of the long-term monitoring in soft X-rays and the first dedicated observations in the hard X-rays of the young XRP \src during its periastron passage in October 2019. The source was found in a bright state on October 22 that coincides with the predictions based on the orbital ephemerides of \cite{2019ATel12890....1S} derived from the optical photometry of the  counterpart. The main results can be summarized as follows:
\begin{enumerate}
\item the first broadband spectrum of the source has been obtained, which turned out to be typical for accreting pulsars and revealed no features which could be associated with cyclotron lines;
\item the observed pulse profiles show little variations with energy, however, comparison of {\it NuSTAR} and {\it XMM-Newton} data suggests that there is some evolution with flux;
\item the pulsar spin period had decreased by $\sim$10\% from the previous measurements five years before;
\item soon after the peak of the outburst the accretion rate unexpectedly dropped by an order of magnitude reaching the lowest value ($\sim10^{36}$~\lum) typically observed at the end of the orbital cycle;
\item during an observation about 20 days later, the source re-appeared at roughly the same luminosity and started to decrease its luminosity in the following observation.
\end{enumerate}

The reason for such a strong spin-up of \src and its timing are unknown. A  natural explanation  would be that we missed an outburst activity of the source which led to the transfer of a large amount of angular momentum to the NS. However, the absence of high quality monitoring X-ray observations covering the time span between 2014 and 2019 imply that this hypothesis cannot be verified based on available data. Here we only note that despite the large absolute spin-up magnitude, the average spin-up rate of $\sim$0.05\,s\,d$^{-1}$ is, in fact, lower than the largest values reported by \cite{2018MNRAS.475.2809G}, which were however, observed over shorter periods of time. Nevertheless, the long-term spin-up in apparent quiescence is very intriguing and is definitely worth further investigation and requires continued monitoring of the source.

Regardless of the physical origin of the spin-up, however, it was accompanied by an even more puzzling behaviour of \src during the current outburst. Indeed, soon after our \nustar observation had been completed, we discovered that the source flux dropped by about a factor of 10 for several subsequent XRT observations. 
 The origin of this drop is unclear, with either an occultation or absorption event or an intrinsic drop of the mass accretion rate being the most obvious options.
Considering the relatively slow decay of the accretion rate during the previous outburst and rapid restoration to typically observed levels the former option appears, however, to be more plausible. Indeed, the $N_{\rm H}$ required to cause an order of magnitude drop of the flux in the 0.3--10\,keV range is around $10^{24}$~cm$^{-2}$ (assuming the intrinsic source spectrum is the same as in the peak of the outburst). Although rather high for XRPs, examples of BeXRPs exhibiting similar behaviour are known in the literature. For instance, SXP~5.05 in the SMC during the detailed monitoring of its outburst in 2013 with {\it Swift}/XRT also revealed strong dips in the X-ray light curve appearing soon after the periastron passage \citep{2015MNRAS.447.2387C}. The observed absorption column $N_{\rm H}$ reached values as high as few $10^{23}$~cm$^{-2}$.
The authors interpreted this behaviour as an occultation of the NS by the circumstellar disc around the Be star that is perpendicular to the orbital plane \citep{2015MNRAS.447.2387C,2019MNRAS.486.3078B}, and a similar scenario could be realized for \src. It is worth to note that the similarity between the behaviour of these two sources is not only in the X-ray light curves. Particularly, both sources show significant variability in the near infrared magnitudes clearly associated with the corresponding X-ray flux drops \citep[see $I$ band light curves in][]{2015MNRAS.447.2387C,2020ATel13426....1S}. 

Alternatively, the observed flux variations could be associated with intrinsic
variations of the accretion rate. Such variability is not uncommon among XRPs,
however, in this case it is unclear why the behaviour of \src changed compared
to that during the previous orbital cycle. We note that besides possible
changes in the structure of the circumstellar disc of the primary, the stability of the
accretion can be affected by an interaction of the accretion flow with the
magnetosphere. Indeed, as the main difference between the two cycles appears to
be the significantly shortened spin period. This suggests that the source must
be relatively close to spin equilibrium and thus the radius of the
magnetosphere must be close to the co-rotation radius. One could, therefore,
anticipate accretion to become unstable at low rates, leading to the onset of the
``propeller'' effect in the current cycle
\citep{1975A&A....39..185I,2016A&A...593A..16T}. The available statistics does
not allow to verify whether the accretion indeed switched off completely at
some stage during the dip and to estimate the  corresponding propeller luminosity.
However, it is interesting to compare the observed flux with the theoretically expected
transitional luminosity. For $B\sim10^{13}-10^{15}$\,G it lies between
$\sim10^{32}-10^{36}$\,erg\,s$^{-1}$, so a rather extreme field of
$\sim10^{15}$\,G would thus be required to explain the observed flux drop as the onset
of a propeller at the observed luminosity. We conclude thus that an obscuration of the source by structures in the wind of the primary or simply wind density fluctuations around the pulsar appear to be a more plausible cause for the observed variability.

\section*{Acknowledgements}
 The authors thank the referee for useful comments and suggestions that helped to improve the paper.
This work was supported by the grant 14.W03.31.0021 of the Ministry of Science and Higher Education of the Russian Federation.
We also acknowledge the support from the Academy of Finland travel grants 324550 (SST), 316932 (AAL), 322779 (JP), 317552 and 331951 (SST, JP), the Vilho, Yrj\"o and Kalle V\"ais\"al\"a Foundation (SST),  the German Academic Exchange Service (DAAD) travel grants 57405000 and 57525212 (VD),  and the  Netherlands Organization for Scientific Research Veni Fellowship (AAM). 
The authors would like to acknowledge networking support by the COST Actions CA16214 and CA16104. We grateful to the \nustar, \xmm and {\it Swift} teams for approving our TOO/DDT proposals.  


\bibliographystyle{aa}
\bibliography{allbib}
\end{document}